\documentclass[twocolumn, letter]{emulateapj}

\usepackage{graphics,graphicx}
\usepackage{fancyheadings}
\usepackage{color}
\usepackage{natbib}

\newcommand{\chandra}{\textit{Chandra}}
\newcommand{\xmm}{\textit{XMM}-Newton}

\def\deg{\hbox{$^\circ$}}

\shorttitle{The impact of the environment on the early stages of radio source evolution}
\shortauthors{Sobolewska et al.}

\begin{document}

\title{The impact of the environment on the early stages of radio source evolution}

\author{Ma{\l}gosia Sobolewska$^1$, Aneta Siemiginowska$^1$, Matteo Guainazzi$^2$, Martin Hardcastle$^3$,
Giulia Migliori$^4$, Luisa Ostorero$^5$, \L ukasz Stawarz$^6$}

\affil{$^1$ Center for Astrophysics $\vert$ Harvard \& Smithsonian, 60 Garden Street, Cambridge, MA 02138, USA}
\affil{$^2$ European Space Research and Technology Centre (ESA/ESTEC),
Kepleriaan 1, 2201 AZ, Noordwijk, The Netherlands}
\affil{$^3$ School of Physics, Astronomy and Mathematics, University of Hertfordshire,
College Lane, Hatfield AL10 9AB, UK}
\affil{$^4$ INAF, Istituto di Radio Astronomia di Bologna, Via P. Gobetti 101, I-40129 Bologna, Italy}
\affil{$^5$ Dipartimento di Fisica - Universit\`a degli Studi di Torino
and Istituto Nazionale di Fisica Nucleare (INFN), Via P. Giuria 1, 10125 Torino, Italy}
\affil{$^6$ Astronomical Observatory, Jagiellonian University, ul. Orla 171, 30-244 Krak\'ow, Poland}

\smallskip
\email{msobolewska@cfa.harvard.edu}


\label{firstpage}

\begin{abstract}
Compact Symmetric Objects (CSOs) 
show radio features such as jets, lobes, hot spots
that are contained within the central 1 kpc region
of their host galaxy. Thus, they are thought to be
among the progenitors of large-scale radio galaxies.
A debate on whether the CSOs are compact
primarily because they are young
or because they are surrounded by a dense medium
impacting their expansion is ongoing.
Until now, attempts to discriminate between the environmental 
and genuine youthfulness scenarios have been inconclusive.
We present a study of three CSOs selected on the 
basis of their puzzling X-ray absorbing properties in prior
Beppo-SAX and/or \chandra\ X-ray Observatory data. Our new \xmm\
observations unambiguously confirm the nature of their X-ray absorbers.
Furthermore, for the first time, our X-ray data reveal the existence
of a population of CSOs with intrinsic hydrogen column density
$N_H>10^{23}$\,cm$^{-2}$ that is different from the population
of X-ray unabsorbed CSOs. The two groups appear to be
separated in the linear size vs. radio power plane. This finding suggests
that a dense medium in X-ray obscured CSOs may be able to confine
the radio jets.
Alternatively, X-ray obscured CSOs could be seen as radio brighter
than their unobscured counterparts either because they reside in a dense
environment or because they have larger jet powers.
Our results help constrain the origin of the X-ray emission and
the location and size of the X-ray obscurer in CSOs,
and indicate that the environment may play a key role during the initial
expansion of a radio source.
\end{abstract}


\section{Introduction}
\label{sec:intro}

Giga-Hertz Peaked-Spectrum (GPS) radio sources are characterized by a 
spectral turn-over at the Giga-Hertz frequencies, and compact
sub-kpc scale radio structures (e.g. review in O'Dea 1998).
A sub-class of Compact Symmetric Objects (CSOs) is identified by 
symmetric double radio structures with the radio intensity
dominated by lobes or hot spots.
These features are also typically observed
in large-scale radio galaxies, but in the CSOs they are contained
within the central 1 kpc region of the host galaxy.
Because by definition CSOs are expected to be observed in the
direction nearly perpendicular
to the radio jet axis, the observed emission is not affected
by beaming. Thus, their linear radio size can be translated into
the source kinematic age if one measures the advance velocity of the 
hot spots.
Multiple studies show that the CSOs radio structures may be
very young, $< 3000$ years old (e.g. Gugliucci et al. 2005;
see also An \& Baan 2012, and references therein).

The reasons for the CSO compactness are still debated.
Fanti et al. (1990) observed an anti-correlation between
the radio turn-over frequency and linear radio size in radio galaxies
with linear sizes $< 20$\,kpc. This finding, along with CSO radio
morphologies resembling miniature versions of those seen in the FRI/FRII
galaxies (Fanaroff \& Riley 1974)
led to the formulation of an evolutionary scenario
in which CSOs are the young counterparts of large scale radio sources
(e.g. O'Dea 1998; Snellen et al. 2003; Kunert-Bajraszewska
et al. 2010). However, there is also increasing observational
evidence in favor of the confinement scenario, in which the CSOs
cannot expand freely because they reside in a dense environment
(e.g. van Breugel 1984; Bicknell et al. 1997; Callingham et al. 2015;
Tingay et al. 2015). In particular, recent radio monitoring of CSOs
provide observational evidence in favor of the existence of
clouds with a complex density distribution in some sources
(Tingay et al. 2015; Callingham et al. 2015, 2017).

\begin{table*}[t]
{\scriptsize
\noindent
\caption[]{\label{tab:data} Log of \xmm\ observations.}
\begin{center}
\begin{tabular}{lllllcl}
\hline\hline
 \# & Source & RA & DEC & Date & Exposure$^a$   \\
    &        &    &     &      & PN/ MOS1/ MOS2 \\
    &        &    &     &      &  [ks]          \\
\hline
1 & 2021+614 & 20 22 06.682 & $+$61 36 58.79 & 2016-05-25 & 24.5/ 30.6/ 36.0 \\
2 & 1934$-$63  & 19 39 25.027 & $-$63 42 45.61 & 2017-04-01 & 22.8/ 27.3/ 27.9 \\
3 & 1946+708 & 19 45 53.520 & $+$70 55 48.72 & 2016-10-21 & 18.1/ 24.9/ 26.3 \\
\hline\hline
\end{tabular}

\noindent
NOTES: $^a$ Effective exposure after applying the standard filtering criteria.
\end{center}
}
\end{table*}

\begin{table*}
\noindent
\caption[]{\label{tab:params} Parameters of the best fitting spectral models}
\begin{center}
\scriptsize
\begin{tabular}{lcccccc}
\hline\hline
Parameter & Unit
& 1934-63
& 1946+708 
& \multicolumn{3}{c}{2021+614}\\

(1) & (2)
& (3)
& (4)
& (5)
& (6)
& (7) \\

 & & {\tt powerlaw} & {\tt powerlaw} & {\tt pexmon} & {\tt mytorus} & {\tt borus02}\\

\hline\hline

Redshift & 
& 0.183
& 0.101 
& 0.227
& 0.227
& 0.227 \\

$N^{\rm gal}_H$ & $10^{20}$\,cm$^{-2}$
& 6.16
& 8.57
& 14.01
& 14.01
& 14.01 \\

\hline

$N^1_H$ & $10^{21}$\,cm$^{-2}$
& $1.2^{+0.8}_{-0.4}$
& $16^{+5}_{-3}$
& $4^{+2}_{-2}$
& $4^{+3}_{-1}$
& $4.2^{+0.6}_{-1.0}$ \\

$N^2_H$ & $10^{23}$\,cm$^{-2}$ 
& $\dots$
& $\dots$
& $3.4^{+0.3}_{-0.3}$
& $3.5^{+0.9}_{-0.8}$
& $3.6^{+0.5}_{-0.6}$ \\

\hline

$\Gamma$ &
& $2.07^{+0.20}_{-0.07}$
& $1.1^{+0.3}_{-0.1}$
& $1.7^{+0.4}_{-0.2}$
& $1.7^{+0.2}_{-0.2}$ 
& $1.7^{+0.2}_{-0.1}$ \\

$\Omega/2\pi$ & 
& $\dots$
& $\dots$
& 1(f)
& $\dots$
& $\dots$ \\

$CF_{\rm tor}$ & 
& $\dots$
& $\dots$
& $\dots$
& 0.5(f)
& $0.6^{+0.2}_{-0.4}$ \\

$C_{\rm pl}$ & 
& $\dots$
& $\dots$
& $0.12^{+0.02}_{-0.06}$
& $0.12^{+0.03}_{-0.05}$
& $0.08^{+0.02}_{-0.03}$ \\

$E_{\rm Fe}$ & keV
& $\dots$
& $\dots$ 
& 6.4(f)
& 6.4(f)
& 6.4(f) \\

$EW^a$ & keV
& $<0.3$
& $<0.2$
& $\dots$
& $\sim$0.2
& $\sim$0.2 \\

\hline

$F_{\rm 2-10~(o)}^b$ & $10^{-13}$\,erg\,s$^{-1}$\,cm$^{-2}$
& $0.8\pm0.1$
& $4.4^{+0.2}_{-0.4}$
& $3.4^{+0.1}_{-0.4}$
& $3.4^{+0.3}_{-0.9}$
& $3.33^{+0.02}_{-0.34}$ \\

$L_{\rm 2-10~(r)}^c$ & $10^{43}$\,erg\,s$^{-1}$
& 0.6
& 1.2
& 10.8
& 11.2
& 9.5 \\

\hline

$\chi^2$ & 
& 233
& 38
& 76
& 73
& 72 \\

d.o.f.$^d$ & 
& 269
& 51
& 69
& 68
& 68 \\

\hline\hline

\end{tabular}
\end{center}
NOTES: Errors represent 1$\sigma$ confidence intervals, except for $EW$ and flux
for which we give 90\% confidence interval uncertainties.
The cutoff energy was fixed at 100, 500, and 300\,keV in models
presented in columns (5), (6) and (7), respectively, but this has
no impact on modeling the $< 10$\,keV \xmm\ data.
$^a$ Equivalent width of the Fe\,K$\alpha$ emission line (PN) computed with respect
to the total continuum.\\
$^b$ Observed (not corrected for absorption) 2--10\,keV (observed frame) flux (PN instrument).\\
$^c$ Intrinsic (corrected for absorption) 2--10\,keV (rest frame) luminosity of the direct power law
(PN instrument).\\
$^d$ Degrees of freedom.\\
\end{table*}

\begin{table*}[t]
{\scriptsize
\noindent
\caption[]{\label{tab:cso} X-ray observed CSO sources}
\begin{center}
\begin{tabular}{rcrccccccc}
\hline\hline
\# & 
Source &
z &
Size &
$\log {\rm L(5GHz)}$ &
Ref &
NH(z) &
$\Gamma$ &
X-ray &
Ref
\\
 &
 &
 &
pc &
W\,m$^{-2}$ &
(radio) &
$10^{22}$\,cm$^{-2}$ &
 &
Observatory &
(X-ray)
\\
(1) & (2) & (3) & (4) & (5) & (6) & (7) & (8) & (9) & (10) \\
\hline
1 &
0035+227 &
0.096 &
31.3 &
24.75 &
[1] &
$1.4^{+0.8}_{-0.6}$ &
(1.7) &
\chandra &
[3]
\\
2 &
0108+338 &
0.668 &
39.0  &
27.33 &
[1] &
$57 \pm 20$ &
(1.75) &
\xmm &
[4]
\\
3 & 
0710+439 &
0.518 &
146 &
27.16 &
[1] &
$1.02^{+0.29}_{-0.22}$ &
$1.75^{+0.11}_{-0.10}$ &
\chandra &
[3]
\\
4 &
1031+567 &
0.460 &
181.8 &
26.98 &
[1] &
$0.50 \pm 0.18$ &
(1.75) &
\xmm &
[4]
\\
5 &
OQ+208 &
0.077 &
11.1 &
25.55 &
[1] &
44-130 &
$1.448^{+0.128}_{-0.002}$ &
\xmm/\chandra/NuSTAR &
[5]
\\
6 &
1511+0518$^a$ &
0.084 &
7.3 &
24.98 &
[1] &
$38^{+40}_{-13}$ &
$3.8^{+0.3}_{-0.4}$ &
\chandra &
[3]
\\
7 &
1607+26$^b$ &
0.473 &
290 &
27.18 &
[1] &
$< 0.18$ &
$1.4 \pm  0.1$ &
\chandra &
[3]
\\
8 &
1718-649 &
0.014 &
2.8 &
24.34 &
[1] &
$0.31 - 0.69$ &
$1.75^{+0.10}_{-0.09}$ &
\chandra/\xmm &
[3, 6]
\\
9 &
1843+356 &
0.764 &
31 &
27.25 &
[1] &
$0.8^{+0.9}_{-0.7}$ &
(1.7) &
\chandra &
[3]
\\
10 &
1934-63$^c$ &
0.181 &
128 &
26.79 &
[1] &
$0.12^{+0.08}_{-0.04}$ &
$2.07^{+0.20}_{-0.07}$ &
\chandra/\xmm &
This work
\\
11 &
1943+546 &
0.263 &
165 &
26.29 &
[1] &
$1.1 \pm 0.7$ &
(1.7) &
\chandra &
[3]
\\
12 &
1946+708$^c$ &
0.101 &
57 &
25.21 &
[1] &
$1.6^{+0.5}_{-0.3}$ &
$1.1^{+0.3}_{-0.1}$ &
\chandra/\xmm &
This work
\\
13 &
2021+614 &
0.227 &
24.5 &
26.54 &
[1] &
$36^{+5}_{-6}$ &
$1.7^{+0.2}_{-0.1}$ &
\chandra/\xmm &
This work
\\
14 &
2351+495 &
0.238 &
179.2 &
26.41 &
[1] &
$0.66 \pm 0.27$ &
(1.75) &
\xmm &
[4]
\\
15 &
0029+3456 &
0.517 &
200 &
27.05 &
[1] &
$1.0^{+0.5}_{-0.4}$ &
$1.43^{+0.20}_{-0.19}$ &
\xmm &
[7]
\\
16 &
0048+3157 &
0.015 &
1.2 &
23.61 &
[1] &
$10.47^{+0.68}_{-1.00}$ &
$1.700^{+0.007}_{-0.006}$ &
\xmm &
[8]
\\
17 &
0431+2037 &
0.219 &
145.6 &
26.47 &
[1] &
$< 0.69$ &
$0.63-2.62$ &
\xmm &
[9]
\\
18 &
0503+0203 (1)&
0.584 &
66.5 &
27.45 &
[1] &
$1.0^{+1.1}_{-0.9}$ &
$2.0^{+0.7}_{-0.5}$ &
\xmm &
[7]
\\
 &
0503+0203 (2)&
0.584 &
66.5 &
27.37 &
[1] &
$0.5^{+0.3}_{-0.2}$ &
$1.62^{+0.21}_{-0.19}$ &
\xmm &
[7]
\\
19 &
1148+5254 &
1.632 &
4.3 &
21.88 &
[1] &
$ < 0.73$ &
$1.58 \pm 0.12$ &
\xmm &
[10]
\\
20 &
J12201+2916 &
0.002 &
1.5 &
21.88 &
[2] &
$< 0.0678$ &
$2.02 - 2.38$ &
\chandra/\xmm &
[11]
\\
21 &
J12342+4753 &
0.372 &
80.8 &
26.09 &
[2] &
$0.06^{+0.12}_{-0.06}$ &
$1.80^{+0.24}_{-0.20}$ &
\chandra &
[12]
\\
22 &
1256+5652 &
0.042 &
56.0 &
24.24 &
[1] &
$8.35^{+4.02}_{-4.64}$ &
$1.57^{+0.14}_{-0.30}$ &
\xmm &
[8]
\\
23 &
J13262+3154 &
0.368 &
285.4 &
26.95 &
[2] &
$0.12^{+0.06}_{-0.05}$ &
$1.74 \pm 0.20$ &
\xmm &
[9] 
\\
24 &
1400+6210 &
0.431 &
218 &
26.80 &
[9] &
$2.9^{+2.0}_{-1.0}$ &
$1.24 \pm 0.17$ &
\xmm &
[9]
\\
\hline\hline
\end{tabular}
\end{center}
NOTES: $^a$ Compton thick candidate (S16).
$^b$ reported to be Compton thick in \xmm\ data (Tengstrandt et al. 2009). Much better quality Chandra data
revealed that the source is unabsorbed. $^c$ reported to be CT in Beppo-SAX data (Risaliti et al. 2003).
Much better quality Chandra data revealed that the source is unabsorbed.References: [1] References in
An \& Baan 2012, [2] Tremblay et al. (2016), [3] S16, [4] Vink et al. (2006), [5] Sobolewska et al. (2018), [6] Beuchert et al. 
(2018), [7] Guainazzi et al. (2006), [8] Singh et al. (2011), [9] Tengstrandt et al. (2009),
[10] Young, Elvis \& Risaliti (2009), [11] Younes et al. (2010), [12] Green et al. (2009).
}
\end{table*}

X-ray observations have the potential to resolve this puzzle because 
the host galaxy environment can be studied in the X-ray band  
via direct measurements of both the interstellar medium (ISM)
temperature and the total equivalent
hydrogen column density, $N_H$, along the line of sight (see Siemiginowska
et al. 2008; Siemiginowska 2009; Tengstrand et al. 2009
for reviews on X-ray emission of GPS sources).
In Siemiginowska et al. (2016; hereafter S16) we presented
the first X-ray study of a sample that included all CSOs
with redshift $z<1$ and measured kinematic ages known by 2010
(16 sources; 6 of them we observed with
\chandra\footnote{\chandra\ X-ray Observatory} for the first time in X-rays).
The majority of the observations were short, but indicated a diverse environment
and moderate column densities. Radio surveys continue providing new CSO kinematic
age estimates, and new CSOs and CSO candidate identifications
(e.g. An \& Baan 2012; Tremblay et al. 2016).

S16 found that the majority of CSOs observed in X-rays 
show moderate or little intrinsic X-ray absorption
(i.e. hydrogen equivalent column density of $N_H < 10^{22}$\,cm$^{-2}$).
However, observations of a few sources suggested that they have
a complex X-ray spectrum with a heterogeneous absorbing medium
and required high quality data for detailed studies
of their emission and absorption properties.
In particular, the X-ray spectrum of OQ$+$208 observed with
\xmm\ (Guainazzi et al. 2004), and \chandra\ and NuSTAR
(Sobolewska et al. 2018)
is dominated by emission reflected from neutral matter. The direct power law
emission is intrinsically absorbed with a column density of
$N_H = (44-130) \times 10^{23}$\,cm$^{-2}$ (Sobolewska et al. 2018).
Thus, OQ$+$208 resembles
an archetypal highly obscured, possibly Compton thick AGN.

Vink et al. (2006) reported significant amount of intrinsic X-ray absorption in
the \xmm\ spectrum of another CSO, 0108$+$388
($N_H = 5.7 \pm 2.0 \times 10^{23}$\,cm$^{-2}$ for
the photon index fixed at $\Gamma = 1.75$).
Singh, Shastri \& Risaliti (2011) found that \xmm\ data
of 0048$+$3157 required three intrinsic absorbers with
a range of hydrogen column densities; in particular,
a partial coverer with $N_H \sim 10^{23}$\,cm$^{-2}$.
Risaliti et al. (2003) reported a Compton thick absorption
in two CSO sources studied with Beppo-SAX: 1934$-$63 and 1946$+$708.
However, these two sources appeared Compton thin in the \chandra\ observations
of S16. Similarly, 1607$+$26 was classified as a Compton thick source in
an \xmm\ observation with  poor S/N ratio (Tengstrand et al. 2009), while S16 found
$N_H < 1.8 \times 10^{21}$\,cm$^{-2}$ in a deeper 40\,ks \chandra\ 
observation.
Finally, in S16 we reported a photon index of $\Gamma \sim 0.8$--1.0 in
1511$+$0518 and 2021$+$614, which is unusually hard for non-beamed sources.
Both sources were flagged as Compton thick CSO candidates.

We observed with \xmm\ the two CSOs with prior Beppo-SAX and \chandra\ 
data (1934$-$63 and 1946$+$708) and one of the two Compton
thick candidates identified in S16 (2021$+$614). Our goal was to constrain 
the presence of any line emission attributable to Fe\,K$\alpha$ originating 
from the locations of these three sources, constrain their X-ray absorption 
properties, and study any underlying correlations between the 
radio size, radio power, and CSO environment.

In this paper we present the results of our new \xmm\ observations.
We discuss the current status of X-ray absorption measurements in
the S16 CSO sample, and the implications for the environment
and initial expansion of radio sources.

\begin{figure}
\includegraphics[width=\columnwidth]{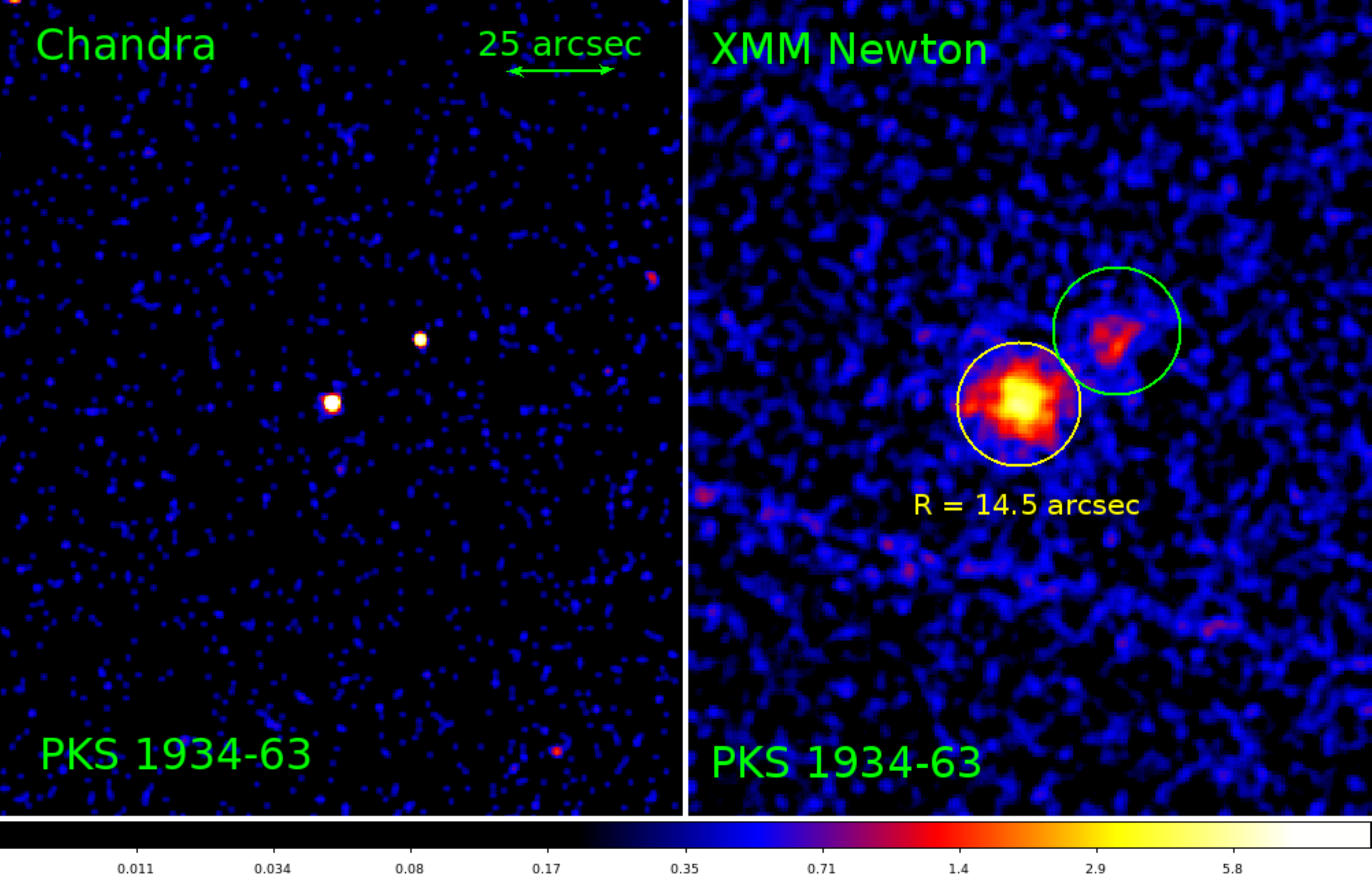}
\includegraphics[width=\columnwidth]{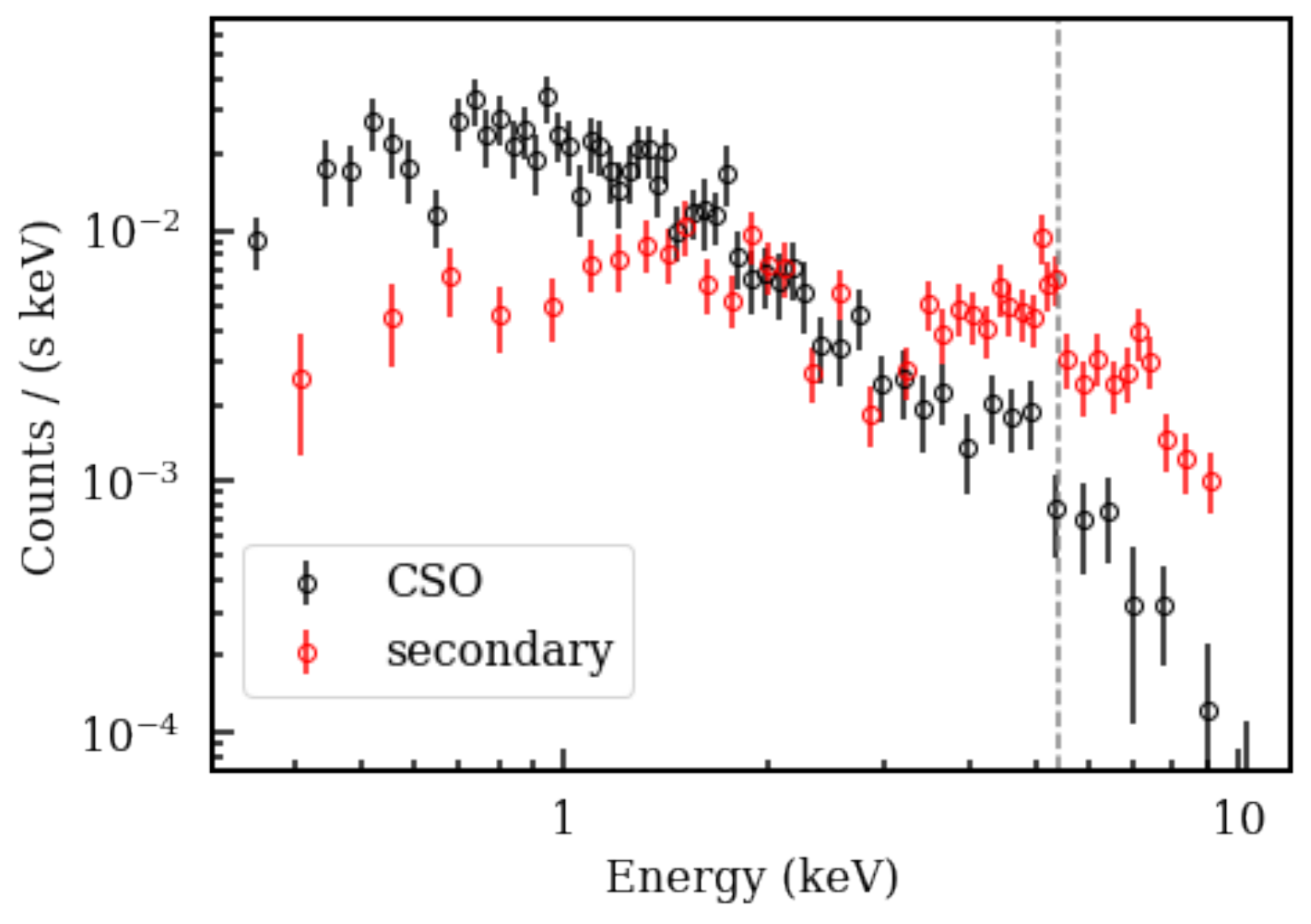}
\caption{Top: \chandra\ ACIS-S (top/left) and \xmm\ EPIC PN (top/right)
images centered on 1934-63.
Yellow and green circles mark the \xmm\ extraction regions
for 1934-63 and the serendipitous source, respectively.
The \chandra\ image is smoothed with a Gaussian kernel with 1.5\arcsec\ radius. 
The \xmm\ image is binned by 16 pixels and smoothed with Gaussian kernel with
the radius of 2.4 \arcsec. The logarithmic color scale corresponds to
the \xmm\ image.
Bottom: \xmm\ PN data corresponding to 1934-63 (black) and the
secondary source (red). The vertical line marks the energy of the Fe\,K$\alpha$
line redshifted to the frame of 1934-63.}
\label{fig:1934}
\end{figure}

\section{Observations}
\label{sec:data}

Table~\ref{tab:data} contains the details of our \xmm\ observations. We present
the data obtained with the EPIC instruments. We used the standard \xmm\ pipeline
products (PPS event lists processed with SAS v.15.0 for  2021$+$614 and 1946$+$708, and SAS v.16.0
for 1934$-$63). We applied the standard screening of the
events, excluding the periods of flaring particle background. The resulted clean 
exposure times for each source are listed in Table~\ref{tab:data}. We note that the high 
background rates impacted approximately half of the originally scheduled exposure of 54\,ks
for 2021$+$614.

Our previous 20\,ks \chandra\ observation revealed a background X-ray source 25\arcsec\ away
from 1934$-$63 in the NW direction (Fig.~\ref{fig:1934}, top/left; S16). This source is also 
visible in our \xmm\ image. Thus, we chose 14.5\arcsec\ and 15\arcsec\ extraction regions for
1934$-$63 and the secondary source, respectively, to avoid mutual flux contamination. There are 
no contaminating sources in the other two CSOs and the standard circular regions with the radius 
of 24\arcsec\ were used for extracting spectra of 2021$+$614 and 1946$+$708. 

We fitted simultaneously the EPIC PN, MOS1 and MOS2 spectra using
{\tt XSPEC} v12.9.1s (Arnaud 1996).

\begin{figure}
\begin{center}
\includegraphics[width=0.9 \columnwidth]{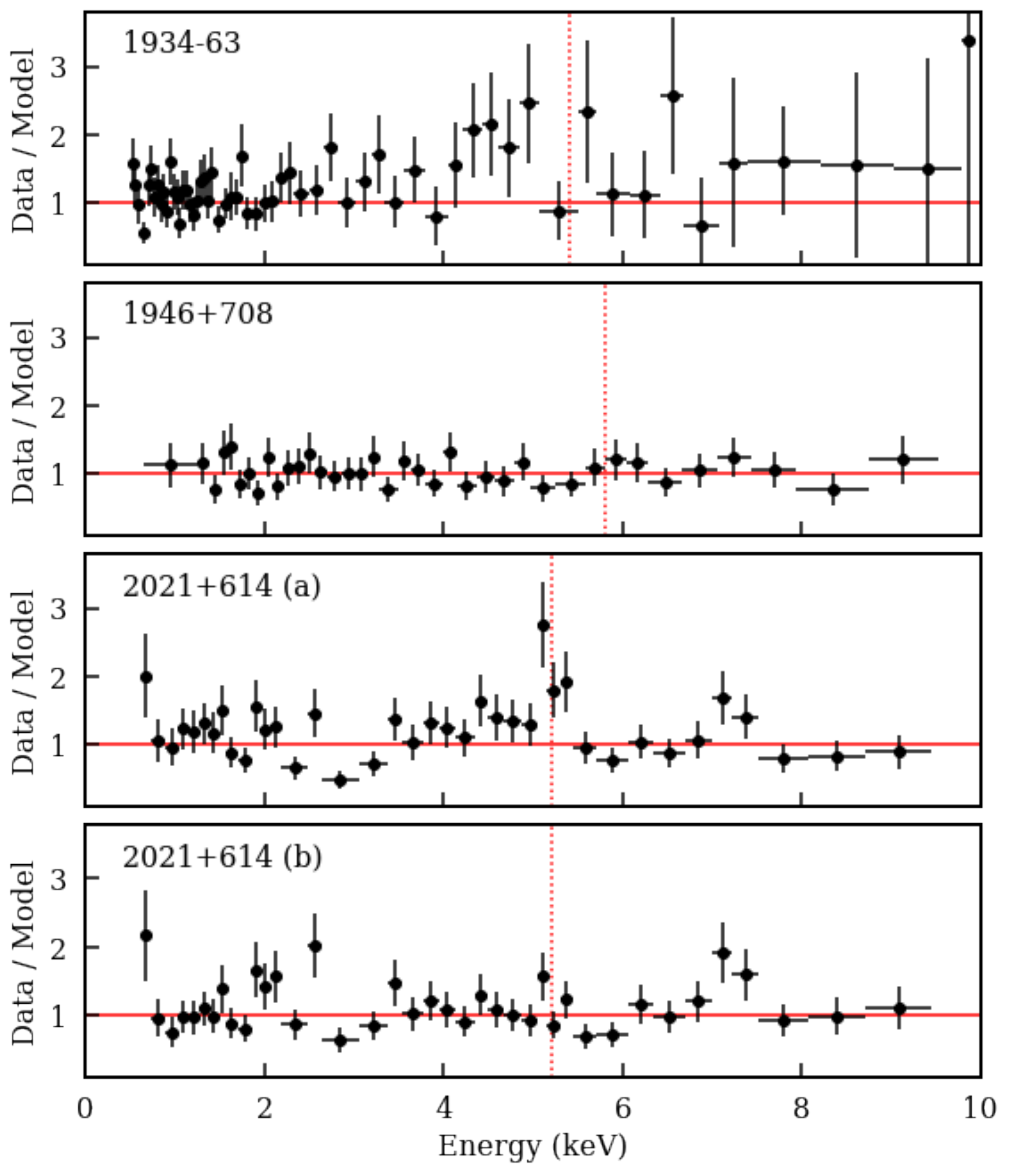}
\caption{Ratios of the PN data to model. Top to bottom:
1934-63, power law modified with Galactic and intrinsic absorbing columns, Tab.~\ref{tab:params}, col. (3);
1946+708, power law modified with Galactic and intrinsic absorbing columns, Tab.~\ref{tab:params}, col. (4);
2021+614 (a), power law with Galactic and intrinsic absorbing columns, with no line emission;
2021+614 (b), the torus model by Balokovi\'c et al. (2018), see Tab.~\ref{tab:params}, col. (7);
models presented for 2021+614 in Tab.~\ref{tab:params} columns (5) and (6) have data/model ratios
that match closely those of 2021+614 (b). Vertical lines mark the rest frame energy of the 
FeK$\alpha$ emission detected only in 2021+614.}
\label{fig:ratio}
\end{center}
\end{figure}

\section{Results}
\label{sec:res}

\begin{figure*}
\begin{center}
\includegraphics[width=17cm]{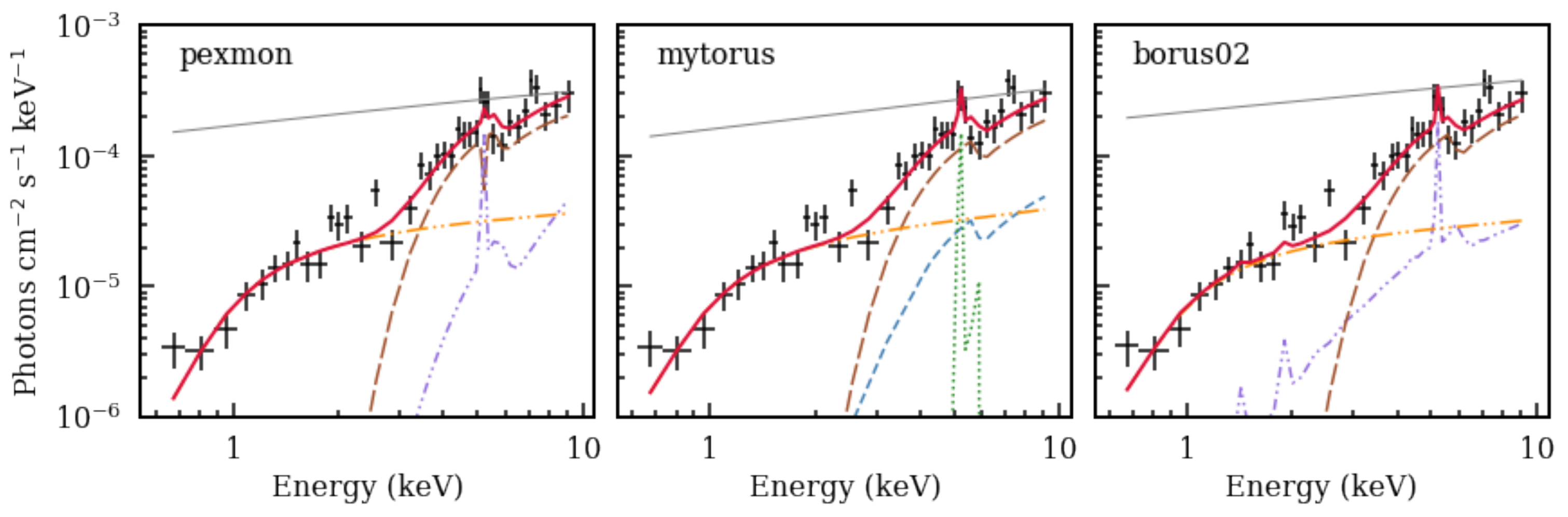}\\
\caption{Best fit 'unfolded' models (red/solid line) and the PN data of 2021$+$614 
(crosses), and model components.
Left: reflection from a slab, Tab.~\ref{tab:params}, col. (5). Middle/Right: 
reflection from a torus, Tab.~\ref{tab:params}, col. (6)/(7).
The model components are as follows: intrinsically absorbed direct continuum
(brown/long-dashed), and scattered direct continuum (orange/dot-long-dashed).
Reflection from a torus and the fluorescent FeK$\alpha$ line emission are computed
and plotted jointly in the case of the {\tt pexmon} and {\tt borus02} models
(purple/dot-short-dashed), and separately in the case of the {\tt mytorus}
model (blue/short-dashed and green/dotted, respectively).
Thin solid gray lines show the de-absorbed direct power law emission.}
\label{fig:efe}
\end{center}
\end{figure*}

Our \xmm\ observations enabled us to refine the intrinsic X-ray absorption
measurements of our targets. We confirm that, in agreement with our prior
\chandra\ observations (S16), the X-ray spectra of 1934$-$63 and 1946$+$708
can be described with a power law model modified by absorption,
$M^1_{\rm abs} = \exp \left(-N^{\rm gal}_H \sigma_E \right) \exp \left( -N^1_H \sigma_{E(1 + z)} \right)$,
where $\sigma_E$ is the photo-electric cross-section and $N^{\rm gal}_H$ and $N^1_H$ are the Galactic and
intrinsic X-ray absorbing columns, respectively ({\tt phabs} and {\tt zphabs} {\tt XSPEC} models).
We detected only a mild intrinsic X-ray absorbing hydrogen column in both sources.
By contrast, we show that 
2021$+$614, a source reported to be a Compton thick
candidate in S16, is indeed absorbed due to a significant intrinsic hydrogen
column density.
Parameters of our best fitting X-ray spectral models are listed in
Table~\ref{tab:params} and the modeling details 
are given below.

\subsection{1934$-$63}

We detected an X-ray source adjacent to 1934$-$63 in both \chandra\ and \xmm\
images (Fig.~\ref{fig:1934}, top). We were not able to model the \chandra\
X-ray spectrum of the secondary source due to a low $S/N$ ratio.

In Fig.~\ref{fig:1934}, bottom, we plot the EPIC PN spectra for the CSO and the secondary
source. This comparison shows that the iron emission originates from
the secondary source, while the \xmm\ spectrum of the CSO is modeled well
with an intrinsically absorbed power law (Tab.~\ref{tab:params}; see
Fig.~\ref{fig:ratio} for the data to model ratio).
Although the \xmm\ data of the CSO appear to contain a hint of iron emission at
$\sim 6.5$\,keV (rest frame), we only find an upper limit on the line's
equivalent width, $EW < 0.3$\,keV (90\% confidence level).
We conclude that the two sources were blended in the Beppo-SAX observation
reported by Risaliti et al. (2003) and this led to the incorrect suggestion that
1934$-$63 may contain a Compton thick obscurer. Here we derived
$N^1_H = 1.2^{+0.8}_{-0.4} \times 10^{21}$\,cm$^{-2}$ for 1934$-$63,
consistent with our \chandra\ result (c.f. S16).

\subsection{1946$+$708}

\xmm\ data are consistent with our prior \chandra\ data (S16) and indicate that
1946$+$708 is not a Compton thick source. No nearby secondary X-ray
sources are seen in our data. The source flux in our observations
is comparable (within a factor of 2) with
the flux reported by Risaliti
et al. (2003) based on the Beppo-SAX data. Thus, we can now rule out a scenario
considered in S16 in which the source switches between the Compton thick and Compton thin states
on time scales of several years. We conclude that, most likely, the marginal detection
of an iron line emission in the Beppo-SAX data that prompted the Compton
thick interpretation was a false positive. In our \xmm\ data
the 90\% confidence level on the equivalent width of the 6.4\,keV line at
the redshift of the source is $EW < 0.2$\,keV.
With the much better quality \xmm\
and \chandra\ data it is evident that the source is only mildly
intrinsically absorbed ($N^1_H = 1.6^{+0.5}_{-0.3} \times
10^{22}$\,cm$^{-2}$; Tab.~\ref{tab:params}; Fig.~\ref{fig:ratio}).

\subsection{2021$+$614}
\label{sec:2021}

Application of an intrinsically absorbed power law model to the \xmm\ spectrum of 2021$+$614
confirmed the unusually hard photon index seen previously in our \chandra\ data (S16).
In addition, we observed residuals near the region corresponding to the rest frame 6.4\,keV
energy, suggestive of the presence of the Fe\,K$\alpha$ fluorescent line emission (Fig.~\ref{fig:ratio};
line emission was not required by our \chandra\ 
data, most likely due to low S/N ratio at energies $>5$\,keV). When we added
a Gaussian line to the model, we obtained 
the rest frame energy of the line $E = 6.3^{+0.2}_{-0.8}$\,keV,
equivalent width $EW = 0.41^{+0.08}_{-0.14}$\,keV,
$\Gamma = 0.34^{+0.05}_{-0.06}$,
and $\chi^2 = 104$ for 69 degrees of freedom.
The apparent hard photon index and the detection of the Fe\,K$\alpha$ line
suggest strong intrinsic obscuration in this source.

To verify this, we constructed a model consisting of an intrinsically absorbed
power law and its reflection from a  neutral matter in the slab geometry
({\tt pexmon}; Nandra et al. 2007) with the Fe\,K$\alpha$ emission computed
self consistently.
However, this model was not able to provide a good fit to the data for
a range of typical AGN photon indices
(e.g. Del Moro et al. 2017; Corral et al. 2011).
Thus, we added a contribution from the direct power law
that is scattered into the line-of-sight, rather than absorbed,
by the obscuring material, and dominates the spectrum at energies
below $\sim$3\,keV.
The total model was given by $M^1_{\rm abs} \times \left[
C_{\rm pl} \times {\tt cutoffpl} + \exp \left( -N^2_H \sigma_{E(1 + z)} \right)
\times {\tt pexmon} \right]$,
with the second intrinsic absorber
modifying only the {\tt pexmon} component. Note that with our choice of
the reflection amplitude, $\Omega/2\pi$, fixed at 1,
the {\tt pexmon} component stands for the combination of the direct
{\tt cutoffpl} and reflected continuum, including the line emission.
The photon indices and normalizations of the scattered {\tt cutoffpl} and
{\tt pexmon} components were linked to each other, and the cut-off energy
was fixed at 100\,keV. We obtained a very good fit with $\chi^2 = 73$ for
67 degrees of freedom, $\Gamma = 1.6 \pm 0.2$, and 
two intrinsic absorbing columns of the order of 
$10^{21}$\,cm$^{-2}$ and $10^{23}$\,cm$^{-2}$
(Fig.~\ref{fig:efe}, left; Tab.~\ref{tab:params}, col. (5)).
The ratio of the scattered and direct power law normalizations was
rather high, $C_{\rm pl} = 0.14^{+0.07}_{-0.06}$ (e.g. Ueda et al. 2007;
Marchesi et al. 2018).

Finally, we explored two models of a toroidal reprocessor ({\tt mytorus}, Yaqoob 2012;
{\tt borus02}, Balokovi\'c et al. 2018). The key difference between
these two models is the flexibility of the {\tt borus02} model to fit for the
torus covering factor, $CF_{\rm tor}$, fixed at 0.5 in the {\tt mytorus} model.
The total model was defined as $M^1_{\rm abs} \times
\left[ C_{\rm pl} \times {\tt cutoffpl} +  M^2_{\rm abs} \times {\tt cutoffpl} +
A \times {\tt torus} \right]$, with $M^2_{\rm abs} = \exp \left[ -N^2_H(\theta) \sigma_{E(1+z)} \right]$
being the zeroth-order continuum given by Yaqoob (2012).
We used the relevant reflected and line emission table models provided by Yaqoob (2012)
and Balokovi\'c et al. (2018) to express the {\tt torus} component.
In both cases the line-of-sight column density was tied to the average toroidal
column density. The inclination angle was fixed at 80$\deg$ since the CSO sources
are believed to be viewed edge on, through the obscuring torus, given
the symmetricity of their radio structures and underlying assumption
that the jet axis is perpendicular to the plane of the torus.
We introduced a normalization constant $A$ following Yaqoob 2012 who argue
that the flexibility in modeling the relative normalizations of the scattered
and line emission is required in order to account for effects such as
a deviation from a perfect toroidal geometry, a range of covering factors,
etc. It was a fit parameter in the {\tt mytorus} model
($A = 2.3\pm1.2$) and it was fixed at 1 (by construction)
in the {\tt borus02} model.
We found good fits for both toroidal models, and the resulting model parameters
were consistent between the two approaches (Fig.~\ref{fig:efe}, middle/right;
Tab.~\ref{tab:params}, cols. (6-7)). Fit parameters of the toroidal models were
consistent with those of the {\tt pexmon} model. The toroidal models result in
the de-absorbed 2--10\,keV rest frame luminosity of the order of
10$^{44}$\,erg\,s$^{-1}$, while the slab geometry model gives 
$\sim6.5 \times 10^{43}$\,erg\,s$^{-1}$.

Models presented in Tab.~\ref{tab:params}, cols. (5--7) are statistically 
equivalent given the current data, and result in the ratio of the data
to model of comparable quality and energy dependence (illustrated in 
the bottom panel in Fig.~\ref{fig:ratio}).
However, all these models imply that the nuclear X-ray emission of 2021$+$614 is
seen through a matter with high intrinsic X-ray absorbing column,
$N^2_H \simeq 3.5 \times 10^{23}$\,cm$^{-2}$, and another
intrinsic absorber with $N^1_H$ of the order of 10$^{21}$\,cm$^{-2}$.

\subsection{Sub-population of X-ray absorbed CSOs}

\begin{figure}
\includegraphics[width=\columnwidth]{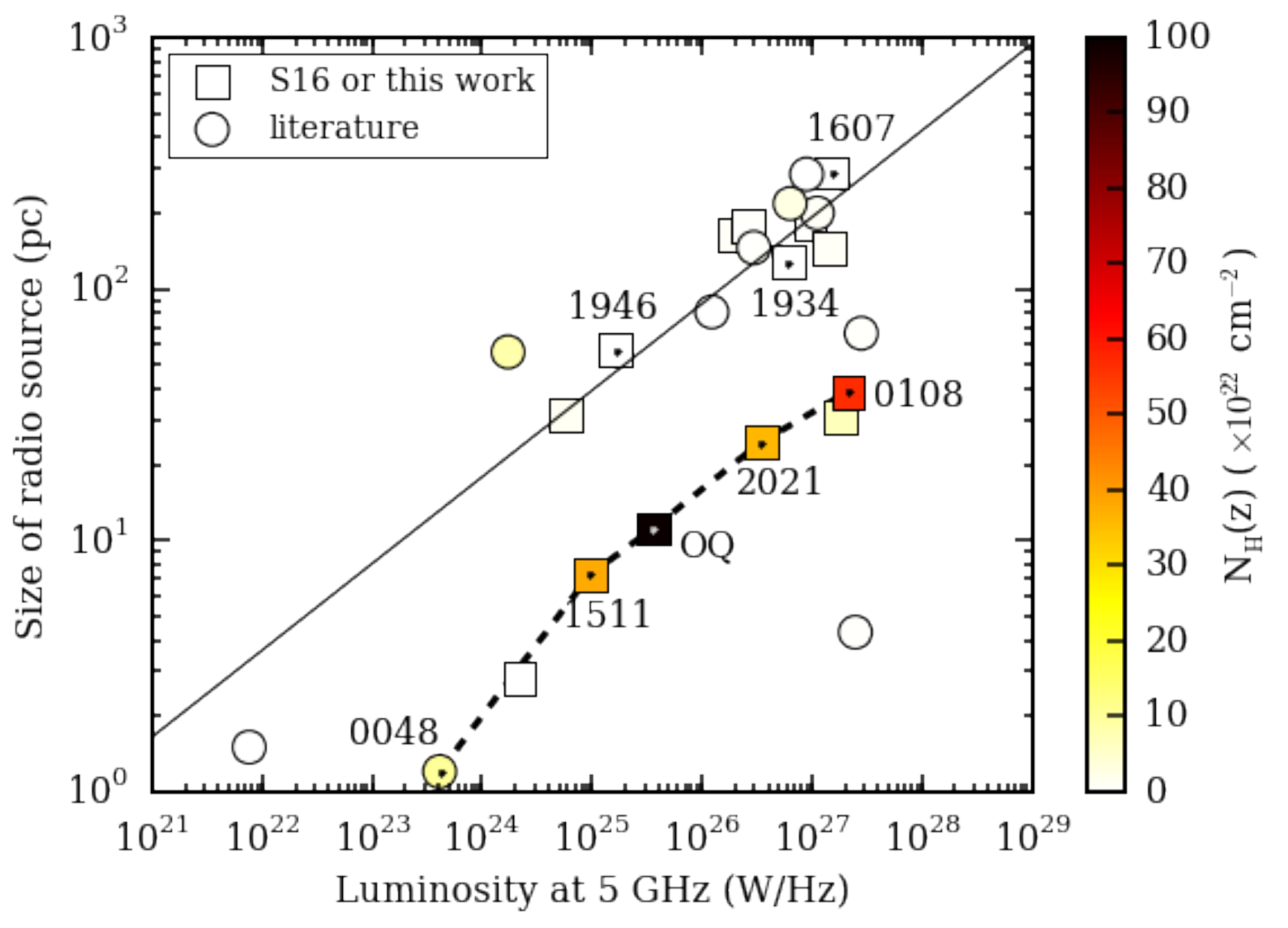}
\caption{Luminosity at 5\,GHz vs. radio source size
for CSOs observed with \chandra\ and/or \xmm. Squares mark
sources with kinematic age measurements studied in X-rays
in S16 and this work. Circles mark
new CSO or candidates (Tremblay et al. 2016; An \& Baan 2012)
with X-ray information collected from the literature 
(Guainazzi et al. 2006; Vink et al. 2006; Green et al. 2009;
Tengstrand et al. 2009; Young, Elvis \& Risaliti 2009;
Younes et al. 2010; Singh, Shastri \& Risaliti\ 2011).
Color coding represents the intrinsic equivalent hydrogen
column density measured from the X-ray spectra. Solid line
marks the relation fitted to the CSO data of Tremblay et al. (2016).
Dashed line connects CSOs with intrinsic $N_H > 10^{23}$\,cm$^{-2}$ (labeled).
Sources labeled as 1946, 1934, 1607 have been previously reported
to be Compton thick but verified to be only mildly absorbed
in our \xmm\ and/or \chandra\ observations (see text for details).}
\label{fig:cso}
\end{figure}

In Tab.~\ref{tab:cso} and Fig.~\ref{fig:cso} we combined
information on the intrinsic X-ray
absorption in CSOs, CSO radio power at 5\,GHz, and radio source
linear size. We stress that sources in Tab.~\ref{tab:cso}
represent all currently known CSOs observed both in
the radio and X-ray bands.
We used the radio data provided by Tremblay et al. (2016), and
for sources from An \& Baan (2012) we derived the 5\,GHz
radio power from their spectral index, $\alpha$, and 4.8\,GHz flux 
as $L_{\rm 5\,GHz} \simeq 4 \pi S_{4.8} D_L^2 / (1+z)^{1 - \alpha}$, where
$D_L$ is the luminosity distance to the source assuming WMAP9 cosmology
(Hinshaw et al. 2013). The X-ray exposures are relatively
deep, with 8 sources observed for $> 30$\,ks, 11 sources observed for
9--30\,ks and only 5 sources observed for 2--5\,ks.

We found that intrinsically X-ray unabsorbed and X-ray absorbed CSOs 
($N_H < 10^{23}$\,cm$^{-2}$ and $N_H > 10^{23}$\,cm$^{-2}$,
respectively) appear to be
separated in the size of the radio source vs. 5\,GHz luminosity plane
(Fig.~\ref{fig:cso}).
In general, the former cluster around the relation
reported by Tremblay et al. (2016; see also Orienti \& Dallacasa 2014),
while the latter are located below this
correlation. To estimate the statistical significance of this
separation we performed two 1-dimensional 2-sample Kolmogorov-Smirnov (K-S)
tests with respect to the CSO radio luminosity and radio size.
The results suggest that while we cannot reject the hypothesis that
the radio luminosities of the two samples come from the same
distribution ($KS = 0.3$ and $p$-${\rm value} = 0.8$), the radio
sizes of the two samples most do not come from
the same distributions at a significance level $> 99$\%
($KS = 0.75$ and $p$-${\rm value} = 9.8 \times 10^{-4}$).
This would support our conclusion that the obscured and unobscured
sources appear to have different radio sizes.

On the other hand, a 2-dimensional 2-sample K-S test should be used in the statistical verification of our results
(e.g. Peacock 1983; Frasano \& Franceschini 1987). However, this test requires that the size of each sample is larger than $\sim 10$.
This condition is not met by our current data as we have only 5 obscured
CSOs. The 2-dimensional 2-sample K-S test can be successfully performed
in the future, when X-ray data become available for an extended CSO sample.
Nevertheless, for our
current unabsorbed and absorbed samples the 2-dimensional 2-sample K-S
test results in $KS = 0.8$ and $p{\rm -value} = 0.02$, which, if confirmed
with future data, would suggest that at 98\% confidence level
the low- and high-$N_H$ CSOs occupy different regions in the radio size
vs. radio luminosity plane.

We conclude that our data are suggestive of the first observational evidence
in the X-ray band for a statistically significant separation in the CSO
population with respect to the density of the CSO environment, as
measured through the intrinsic equivalent hydrogen column density,
and radio properties such as the source size, and perhaps also source
radio luminosity.

\section{Discussion}

We observed three Compton thick CSO candidates with \xmm\ and
were able to refute the presence of a Compton thick
absorber in 1934$-$63 and 1946$+$708, and detect a significant
intrinsic X-ray absorbing column in 2021$+$614.
The X-ray spectrum of 2021$+$614 shows that the intrinsic X-ray
power law is strongly attenuated and it
is seen primarily in a reflection from a neutral matter.
In particular, the detection
of the 6.4 keV iron emission line accompanying the reflection continuum
indicates that 2021$+$614 is not a beamed source (c.f. Giommi et al. 2007
and D'Abrusco et al. 2014 who classified the source as a blazar).
Our findings are in agreement with the results of Barter et al. (1984) who reported
on strong reddening in the optical spectroscopic observations of 2021$+$614
performed with the Multiple Mirror Telescope.

We note that the best-fit photon indexes of the power law models
applied to the X-ray spectra of these three CSOs cover a range of
$\Gamma \sim 1$--2 (Tab.~\ref{tab:params}).
Thus, it is likely that different physical processes give rise to the
intrinsic X-ray emission in these sources (e.g. inverse-Compton processes
in expanding radio lobes, X-ray jet, X-ray corona).

We observed that the X-ray absorbed and unabsorbed CSOs appear to be
separated in the radio size vs. 5\,GHz radio luminosity plane (Fig.~\ref{fig:cso}).
We performed 1-dimensional 2-sample K-S tests with repspect to the radio
properties, and found that the radio sizes of the absorbed and unabsorbed
CSO samples could form two populations. In addition, the
results of a 2-dimensional 2-sample K-S test confirmed a possible existence
of such two populations. However, we caution that the 2-dimensional 2-sample
K-S test requires a larger number of the absorbed sources (as least twice
as many as in the current sample) to be valid.

Our results could be interpreted in two ways.
It could be that the CSOs with a high intrinsic X-ray absorbing column,
$N_H > 10^{23}$\,cm$^{-2}$; 0048$+$3157, OQ$+$208, 2021$+$614, 1511$+$518,
0108$+$388) have less evolved radio structures (smaller linear
radio size) than X-ray unabsorbed CSOs with the same 5\,GHz radio power.
Thus, the environment in which a radio source
expands might be able to confine the source growth already at the earliest
stages of its evolution. This would provide the first observational evidence
in the X-ray band that some compact jets may be frustrated by a dense
environment and their kinematic ages may be underestimated.

An alternative way of explaining the separation visible in Fig.~\ref{fig:cso}
is that the X-ray obscured sources have higher radio luminosity than the
X-ray unobscured sources for a given radio size. The
two factors that drive radio luminosity for a given 
size of a radio source are environmental density and jet power.
Thus, one possible explanation is that the X-ray obscured sources
are brighter for a given size because they were born in a denser
environment. Another option is that they are brighter than the unobscured CSOs
with the same size because they have higher jet power, which is
related to central density, through e.g. the accretion
rate (e.g. Allen et al. 2006; Tchekhovskoy et al. 2011; Ghisellini et al. 2014),
or directly as in the model of Stawarz et al. (2008).

Even though with the current data it is not possible to distinguish between
these two explanations, we stress that in both cases the density of the CSO
environment plays a key role. Moreover, our results rise a question about
the origins of the X-ray emission and the location of the X-ray obscurer.
If the CSO X-ray emission originates primarily from an active nucleus that becomes
attenuated by obscuring material in the vicinity of the nucleus, then we
should also see heavily obscured CSOs with larger radio structures.
In contrast, the CSOs in our sample show low intrinsic X-ray
absorption for the radio size exceeding a few tens of parsecs.
If the CSO X-ray emission
is dominated by inverse-Compton emission of the radio lobes
(e.g. Stawarz et al. 2008; Ostorero et al. 2010), then the CSOs
would become unobscured once their radio lobes expand beyond the 
region occupied by the dense material.
Thus, it could be that the obscurer is located on scales smaller
than a few tens of parsecs.
Alternatively, the obscuring material could
be destroyed or removed by the expanding radio source.

Recently, Ostorero et al. (2017) addressed the question
of the relative location of the X-ray and radio absorbers
in the CSOs.
They confirmed a positive correlation
between the total absorbing column density ($N_H$, X-rays) and the
neutral hydrogen column ($N_{HI}$, radio), originally found by
Ostorero et al. (2010). The existence
of such a correlation suggests that the gas responsible for the X-ray
and radio absorption may be part of the same, possibly unsettled,
hundred-parsec scale structure.

We note that there are three sources whose X-ray properties place them
in the vicinity of the X-ray absorbed sources in Fig.~\ref{fig:cso} and/or
significantly below the correlation reported by Tremblay et al. (2016).
Yet, they appear to be X-ray unabsorbed with $N_H$ of the order of
$10^{20}-10^{22}$\,cm$^{-2}$. 
We observed one of them with \chandra\ (1843$+$356;
$\rm L_{\rm 5\,GHz} = 1.8\times 10^{27}$\,W\,Hz$^{-1}$, $\rm LS = 31$\,pc)
for 5\,ks and detected only $10.8\pm3.3$ counts.
Thus, the measured intrinsic absorption is highly uncertain (S16).
However, we should be able to robustly constrain both the $\Gamma$ and
intrinsic $N_H$ in this source with the pending \xmm\ observation.
The two remaining sources display characteristics that are not typical
to our CSO X-ray sample.
PKS~1718$-$649 ($\rm L_{\rm 5\,GHz} = 2.2\times 10^{24}$\,W\,Hz$^{-1}$,
$\rm LS = 2$\,pc) is to date the only $\gamma$-ray emitting CSO (Migliori et al. 2016).
While at $z = 1.632$, 1148$+$5254 (a.k.a. SDSS J114856.56+525425.2;
$\rm L_{\rm 5\,GHz} = 2.4\times 10^{27}$\,W\,Hz$^{-1}$,
$\rm LS = 4.3$\,pc) is the only source in Fig.~\ref{fig:cso}
with redshift above 1. Additionally, its radio data indicate a presence
of a larger scale (older) emission not associated with the younger
radio source (Tremblay et al. 2016; Young, Elvis \& Risaliti 2009).

The CSOs constitute the youngest resolved radio structures and provide information
on early stages of the radio source expansion, and its impact on the feedback
between the evolution of a galactic environment and black hole growth. The future X-ray
observations of a large number of the CSOs are necessary for a statistical
verification of the trends observed in this work.

\section{Conclusions}

We studied the X-ray and radio properties of a sample of Compact Symmetric Objects.
The sample included all currently identified
CSOs with (1) X-ray spectral information obtained with \chandra\
and/or \xmm, (2) morphological information on their compact radio structures,
and (3) radio band spectral information.
We considered a tree-dimensional parameter space with axes corresponding to
the radio luminosity at 5\,GHz, $L_{\rm 5\,GHz}$, radio source linear size, $\rm LS$,
and intrinsic X-ray hydrogen column density, $N_H$.
Our main findings are as follows.

\begin{itemize}

\item Two CSOs, 1934$-$63 and 1946$+$708, observed with \xmm\  are only mildly
X-ray obscured indicating that they are not 
Compton-thick sources (dismissing the original claim based on the lower-quality
BeppoSAX data),
while the third CSO, 2021$+$614, is highly absorbed with the X-ray intrinsic column
density of $\sim 3.5 \times 10^{23}$\,cm$^{-2}$.

\item X-ray obscured and unobscured CSOs appear to occupy separate regions
in the ($L_{\rm 5\,GHz}$, LS, $N_H$) parameter space.

\item One interpretation of the above separation is that X-ray absorbed CSOs
have smaller radio sizes than X-ray unabsorbed CSOs. Alternatively, X-ray absorbed
CSOs could be more radio luminous than X-ray unabsorbed CSOs with the same radio size.

\item We argued that in both cases the high density environment plays the key role:
(1) High density ISM could prevent the radio source from expanding freely.
(2) Radio source born in a dense environment would appear more radio luminous than
a source born in low density environment, as a high density environment
is able to sustain higher mass accretion rate onto a black hole than a low density
environment. Consequently, jet power and hence radio luminosity increase with
increasing mass accretion rate.

\item If X-ray emission originates in radio lobes, then our results place an upper
limit on the distance of the
high column density clouds from the center as we did not detect high intrinsic X-ray
obscuration in CSOs with LS exceeding a few tens of parsecs.
\end{itemize}

Expanding our sample of X-ray CSOs is necessary in order to verify
the statistical significance of the separation between the X-ray
obscured and unobscured CSO in the radio power vs. radio size plane
and draw important conclusions on the effect of the environment
on the early evolution of the radio sources.
Deep \xmm\ exposures of 1511$+$518 (Compton thick candidate identified in S16) and 1843$+$356 would allow us to put more stringent constraints
on the intrinsic X-ray absorbers in these sources. Excellent candidates for
a follow-up X-ray study, with radio properties that
place them near the emerging X-ray absorbed branch in Fig.~\ref{fig:cso},
have been recently identified by An \& Baan (2012) and Tremblay et al. (2016).
Ongoing and future radio surveys will keep provide new CSO targets for
X-ray follow-ups.

\acknowledgements
The authors thank G. Risaliti and M. Balokovi{\'c} for discussions regarding
spectral modeling of the studied sources. M.S. and A.S. were supported by NASA
contract NAS8-03060 (Chandra X-ray Center). M.S. acknowledges the Polish NCN
grant OPUS 2014/13/B/ST9/00570. 
{\L} . S. was supported by the Polish NCN grant 2016/22/E/ST9/00061.
This project was funded in part by the Chandra
grant GO4-15099X. The \xmm\ observations were obtained as part of the 078461
project.

\end{document}